\documentclass{ifacconf}

\usepackage{graphicx}      
\usepackage{natbib}        
\usepackage{epstopdf}
\usepackage{epsfig}
\usepackage{amsmath}
\usepackage{algcompatible}
\usepackage{amssymb}
\usepackage[usenames, dvipsnames]{color}
\usepackage{algorithm}
\usepackage{algpseudocode}
\usepackage{mathtools}
\usepackage{bbm}
\newcommand{\REAL}{\ensuremath{\mathbb{R}}}

\newcommand{\beq}{\begin{equation}}
\newcommand{\eeq}{\end{equation}}
\newcommand{\bea}{\begin{eqnarray}}
\newcommand{\eea}{\end{eqnarray}}
\newcommand{\bean}{\begin{eqnarray*}}
\newcommand{\eean}{\end{eqnarray*}}
\newcommand{\bcen}{\begin{center}}
\newcommand{\ecen}{\end{center}}
\newcommand{\bitm}{\begin{itemize}}
\newcommand{\eitm}{\end{itemize}}

\def\sgcnv{\hbox{$\, \bigcirc \,$\kern-0.9em\hbox{\mgop}$\,$}} 
\def\supgeno{\hbox{$\, \bigcirc \,$\kern-1.0em\hbox{$\wedge$}$\,$}} 
\def\infgeno{\hbox{$\, \bigcirc \,$\kern-1.0em\hbox{$\vee$}$\,$}} 
\newcommand{\mgop}{\ensuremath{\star}}  

\newcommand{\vct}[1]{\ensuremath{\boldsymbol{#1}}}  
\newtheorem{definition}{Definition}
\newtheorem{assumption}{Assumption}
\newtheorem{remark}{Remark}

\newcommand{\abs}[1]{\ensuremath{\left\vert #1\right\vert}}
\newcommand{\norm}[1]{\ensuremath{\left\| #1\right\|}}
\newcommand{\paren}[1]{\ensuremath{\left( #1\right)}}
\newcommand{\clint}[1]{\ensuremath{\left[ #1\right]}}

\newcommand{\set}[1]{\ensuremath{\left\{ #1\right\}}}

\newcommand{\matr}[1]{\ensuremath{\clint{\begin{array} #1 \end{array}}}}
\newcommand{\expe}[1]{\ensuremath{\mathbb{E}\set{#1}}}

\DeclareMathOperator{\Tr}{\mathbf{tr}}

\begin{document}
\begin{frontmatter}

\title{
State Estimation with Secrecy against Eavesdroppers}

\thanks[footnoteinfo]{This work was supported by NSF CNS-1505799 grant and the Intel-NSF Partnership for Cyber-Physical Systems Security and Privacy. }

\author{Anastasios Tsiamis, Konstantinos Gatsis, George J. Pappas}

\address{Department of Electrical and Systems Engineering, University of Pennsylvania, 200 South 33rd Street, Philadelphia, PA 19104, United States (e-mails: [atsiamis, kgatsis, pappasg]@seas.upenn.edu).}

\begin{abstract}              
We study the problem of remote state estimation, in the presence of an eavesdropper. An authorized user estimates the state of a linear plant, based on the data received from a sensor, while the data may also be intercepted by the eavesdropper.
To maintain confidentiality with respect to state, we introduce a novel control-theoretic definition of perfect secrecy requiring that the user's expected error remains bounded while the eavesdropper's expected error grows unbounded. 
We propose a secrecy mechanism which guarantees perfect secrecy by randomly withholding sensor information, under the condition that the user's packet reception rate is larger than the eavesdropper's interception rate. Given this mechanism, we also explore the tradeoff between user's utility and confidentiality with respect to the eavesdropper, via an optimization problem. Finally, some examples are studied to provide insights about this tradeoff.
\end{abstract}


\end{frontmatter}

\section{Introduction}

The recent emergence of the Internet of Things as a collection of wirelessly connected sensor and actuator devices has given rise to significant cyber-security concerns~\citep{cardenas2008research, sandberg2015cyberphysical}. Research efforts have targeted, for example, denial-of-service attacks~\citep{amin2009safe,gupta2010optimal}, privacy issues~\citep{LeNyPappas_DP}, as well as data integrity of compromised sensors~\citep{fawzi2014secure, mo2014resilient, pajic2014robustness}. However, the broadcast nature of the wireless medium opens up further vulnerabilities in such connected systems~\citep{Survey_Wireless_Security}. A fundamental vulnerability is confidentiality against \emph{eavesdroppers} who may intercept the transmitted information. This becomes particularly crucial for sensor or actuator data, which convey critical information about the physical system state.

Encryption and cryptography-based tools are commonly employed for confidential communication~\citep{Cryptography}, dating back to the work of~\citep{Shannon_secrecy}. These approaches rely on encrypting communication messages with public or private keys, in order to achieve confidentiality against computationally limited eavesdroppers. These are generic tools, typically employed at intermediate layers of the communication protocol stack, and they do not take into account the characteristics of the application for which confidentiality is required, or the characteristics of the physical layer used for message communication, e.g., the wireless medium.

Additionally it is possible to exploit the physical layer in order to achieve confidentiality, usually termed secrecy in this context~\citep{Wyner_wiretap, LiangPoor2008Secure, Secrecy_MIMO, Proc_IEEE_Special_issue}. This approach models the eavesdropper as overhearing the communication over a channel that is degraded compared to the legitimate channel. This channel disparity may be exploited using information-theoretic tools to achieve a positive communication rate with secrecy. This information-theoretic approach may be applied to problems in remote estimation and control, as discussed in preliminary works \citep{li2011communication, Wiretap_estimation}. 
However, the construction of practical secrecy codes is still an active area of investigation~\citep{Proc_IEEE_Special_issue}.

In this paper, we take an alternative approach by introducing a novel control-theoretic definition of secrecy and by designing simple mechanisms that meet this definition. More specifically, we consider a sensor transmitting the outputs of an unstable linear dynamical system to a legitimate user, while an {eavesdropper} tries to intercept the sent messages compromising confidentiality. Communication follows the packet-based paradigm commonly used in networked control systems~\citep{Sinopoli2004Kalman, hespanha2007survey}, where the user and the eavesdropper respectively receive and intercept packets with different success rates. Our definition of perfect secrecy requires that any state estimate devised by the eavesdropper, suffers from an unbounded expected error in the limit, while the legitimate user still tracks the state with a bounded error (Section~\ref{sec:problem_formulation}).

In Section~\ref{Section_Perfect_Expected_Secrecy}, we show that perfect secrecy is possible as long as the packet success rate of the user is larger than the packet interception rate of the eavesdropper (Theorem~\ref{THR_perfect_secrecy_pc}). In fact, this is achieved by a simple mechanism, which randomly withholds information with the appropriate rate at the sensor. Then, estimation at both the user and the eavesdropper, follows two respective intermittent Kalman filter problems \citep{Sinopoli2004Kalman}. To achieve perfect secrecy, we exploit both  the unstable dynamics and the  inferiority of the eavesdropper's rate. The latter is similar in spirit to the degraded channel assumption in information-theoretic approaches~\citep{Proc_IEEE_Special_issue}. Our approach differs by explicitly considering a control-theoretic definition of secrecy and by subsequently employing control-theoretic mathematical tools. Furthermore, our packet-based communication model yields a simple secrecy mechanism, in contrast to the problem of developing appropriate coding in, e.g., \cite{li2011communication, Wiretap_estimation}.

In Section~\ref{Section_Relaxed}, we relax the perfect secrecy requirement, 
by seeking to achieve the minimum estimation error at the user, as long as the eavesdropper's error is larger than a desired lower bound. Due to the lack of analytical expressions for the intermittent Kalman filter error, we relax the problem by replacing the errors with known upper and lower bounds. The resulting optimization problem can be solved efficiently using bisection, and can be used as a tool for approximate, yet quantitative, analysis of the tradeoff between secrecy and utility. 
In Section~\ref{sec:scalar}, we illustrate this analysis in a scalar system example, which reveals how the parameters of the system influence this tradeoff. We also present numerical results, which demonstrate the effect of employing the proposed secrecy mechanism to a second order system, for a typical sample of packet reception sequences. We conclude with remarks in Section~\ref{sec:conclusion}.

\section{Problem Formulation}\label{sec:problem_formulation}

\begin{figure}[t] \centering{
\includegraphics[scale=0.8]{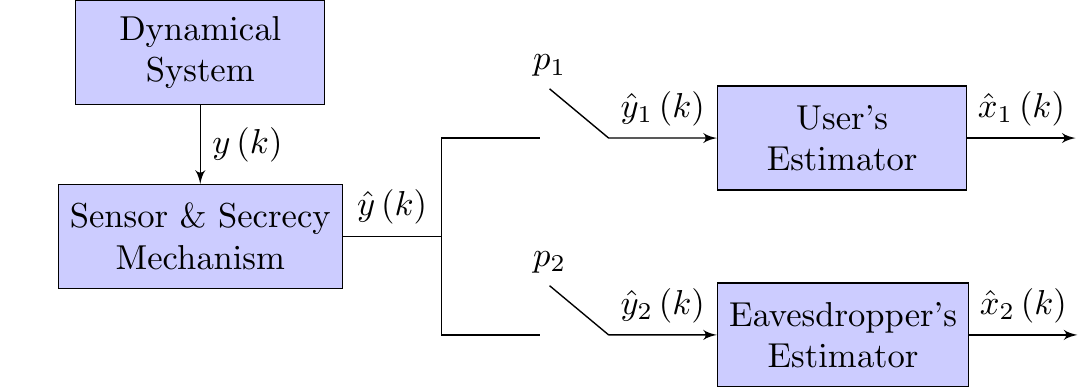}  
\caption { A sensor collects the output $y\paren{k}$ of the dynamical system \eqref{EQN_system}. Equipped with a secrecy mechanism, it produces a distorted version $\hat{y}\paren{k}$ of the output, and sends it to the channel. The user and the eavesdropper receive the distorted output with probability $p_{1}$ and $p_{2}$ respectively, and compute the minimum square error estimates $\hat{x}_{1}\paren{k}$ and $\hat{x}_{2}\paren{k}$.}
\label{Figure_ProblemSetup} }
\end{figure}

The considered remote estimation architecture is shown in Figure \ref{Figure_ProblemSetup}  and consists of a sensor observing a dynamical system, a legitimate user, and an eavesdropper. We 
consider the following linear dynamical system:
\begin{equation}
\begin{aligned} \label{EQN_system}
x\paren{k+1}&=Ax\paren{k}+w\paren{k}\\
y\paren{k}&=Cx\paren{k}+v\paren{k}
\end{aligned}
\end{equation}
where $x\paren{k}\in \REAL^{n}$ is the state, $y\paren{k}\in \REAL^{m}$ is the output and $k\in \mathbb{N}$ is the (discrete) time.  Signals $w\paren{k}\in \REAL^{n}$ and $v\paren{k}\in \REAL^{m}$ are the process and measurement noise respectively. They are modeled as independent Gaussian random variables with zero mean and covariance matrices $Q$ and $R$ respectively.
The initial state $x_{0}$ is also a Gaussian random variable with zero mean and covariance $\Sigma_{0}$. Covariance matrices $R$, $Q$, $\Sigma_{0}$ are assumed to be 
positive definite. In more compact notation $R,Q,\Sigma_{0}\succ 0$, where $\succ$ ($\succeq$) denotes comparison in the positive definite (semidefinite) cone. The system is assumed to be unstable, i.e., its spectral radius $\rho(A) = \max_{i} |\lambda_i(A)|>1$. We also make all assumptions needed for a Kalman filter to be well behaved, namely $\paren{A,Q^{\frac{1}{2}}}$ is stabilizable and $\paren{A,C}$ is detectable.

The sensor communicates over a channel with two outputs/receivers as shown in Figure \ref{Figure_ProblemSetup}. The input to the channel is denoted by $\hat{y}\paren{k}$. The first output, denoted by $\hat{y}_{1}\paren{k}$, is the authorized one to the user, while the second, denoted by $\hat{y}_{2}\paren{k}$, is the unauthorized one to the eavesdropper.
The communication is organized in packets, which consist of sufficiently large number of bits to neglect quantization errors~\citep{Sinopoli2004Kalman, hespanha2007survey, GatsisEtal14}.

Communication with the user is unreliable, i.e., may undergo packet drops. Additionally, communication is not secure against the eavesdropper, i.e., the latter may intercept transmitted packets. Packet drops and packet interceptions are modeled as independent and identically distributed (i.i.d.) over time and across outputs -- see Remark~\ref{remark:iid_model} for further discussion on this model.
In particular, we denote by $p_{1}$ the probability that a packet of the authorized output is received by the user. Similarly, $p_{2}$ is the probability that a packet of the unauthorized output is intercepted. Thus, the outputs of the channel are modeled as follows:
\begin{equation}\label{EQN_channel_model} 
\hat{y}_i\paren{k}=\left\{
\begin{aligned}&\hat{y}(k), \quad \text{with prob. } p_{i}\\
&\varepsilon,  \quad \text{with prob. } 1-p_{i}
\end{aligned}\right.
\end{equation}
for $i=1$, $2$, where
symbol $\varepsilon$, is used to represent the ``no information" outcome.

Our goal is to design a secrecy  mechanism at the sensor, so that communication over the channel conveys sufficient plant state information to the user, but limited state information to the eavesdropper. In particular, the sensor selects what message $\hat{y}(k) \in \mathbb{R}^m\cup\set{\varepsilon}$ is sent over the channel at each time step $k$. The symbol $\varepsilon$ indicates that no information is sent, which we hereby for simplicity do not distinguish from the case of a packet drop in the channel. Formally, a secrecy mechanism $\pi$ dictates how $\hat{y}(k)$ is selected, possibly randomly, given all the available information at the sensor at time $k$. This information includes all past measurements and sent messages, i.e., $y\paren{0}, \ldots, y(k), \hat{y}(0), \ldots, \hat{y}(k-1) $. The mechanism does not know the success of previous packets (values of $\hat{y}_{i}\paren{k}$), in particular, no acknowledgement from the user is assumed.

In this architecture, we assume that all system and noise parameters $A, C, Q$, $R, \Sigma_0$ as well as the probabilities $p_{1}$, $p_{2}$  are public knowledge, available to all involved entities, i.e., the sensor, the user, and the eavesdropper (see also Remark \ref{REM_adversary_knowledge}). Moreover, both the user and the eavesdropper know the mechanism $\pi$.
Under those assumptions, both the user and the eavesdropper are interested in using their received information to obtain a mean square error estimate of the system state.  
Let us denote by $\hat{x}_i(k)$ the mean square error estimate at the user ($i=1$) and the eavesdropper ($i=2$) defined as
\begin{equation}\label{eq:estimate_definition}
\hat{x}_{i}\paren{k}=\expe{ x\paren{k}\vert \vct{\hat{y}}_{i,k}}
\end{equation}
where $\vct{\hat{y}}_{i,k}=\paren{\hat{y}_i\paren{0}, \ldots, \hat{y}_i\paren{k}}$. The corresponding estimation error covariance is given by
\begin{equation}
P_{i}(k\vert k)=\expe{\paren{x\paren{k}-\hat{x}_{i}\paren{k}}\paren{x\paren{k}-\hat{x}_{i}\paren{k}}^{\intercal}\vert \vct{\hat{y}}_{i,k}}
\end{equation}
We also define the covariance of the prediction error as
 \begin{equation}\label{EQN_prediction_covariance}
 P_{i}\paren{k+1}=AP_{i}\paren{k\vert k}A^{\intercal}+Q
 \end{equation} 
 with $P_{i}\paren{0}=\Sigma_{0}$ at time $k=0$. Throughout this work, we quantify the uncertainty about the state using the expected value of the prediction error.

We are now ready to introduce our notion of secrecy, requiring that the eavesdropper's uncertainty grows unbounded, whereas the user's uncertainty remains bounded. We term this \emph{perfect expected secrecy}.

\begin{definition}[Perfect Expected Secrecy]\label{def:perfect_secrecy}
Given a system as in \eqref{EQN_system} and a channel model as in \eqref{EQN_channel_model}, we say that perfect expected secrecy is achieved, under a sensor mechanism $\pi$, if and only if for any initial condition $\Sigma_{0}\succ 0$, both of the following conditions hold:

\begin{align}\label{EQN_Definition_Perfect_Secrecy_1}
\lim_{k\rightarrow \infty}{\Tr{\expe{P_{2}\paren{k}}}}&=\infty\\ \label{EQN_Definition_Perfect_Secrecy_2}
\limsup_{k\rightarrow \infty}{\Tr{\expe{P_{1}\paren{k}}}}&<\infty
\end{align}
where  $\Tr$ is the trace operator.\hfill $\diamond$
\end{definition}

   We point out that this is an asymptotic notion of secrecy as time $k$ grows. Since the system has an initial state with finite variance, it is evident that the eavesdropper can, e.g., maintain a trivial estimate $\hat{x}_2(k) =0$ that has large but nevertheless finite expected prediction error at any finite time step $k$. We also note that the secrecy constraints are required in expectation, not almost surely. By our model, there is always a non-zero probability event that the eavesdropper successfully intercepts a long sequence of messages and intuitively maintains a good state estimate. 

To have a well-posed problem under nominal system operation without any secrecy concerns, we assume that when the sensor sends all output measurements, the user has bounded uncertainty.

\begin{assumption} \label{ASSUM_nominal_error}
If the mechanism $\hat{y}\paren{k}=y\paren{k}$ is employed for all $k \geq 0$, then the user's expected error is bounded
 \begin{equation}\label{EQN_assumption_bounded_nominal_error}
 \limsup_{k\rightarrow \infty}{\Tr{\expe{P_{1}\paren{k}}}}<\infty 
 \end{equation}
for any initial condition $\Sigma_{0}\succ 0$.
\end{assumption}

The goal of the following section is to propose a simple secrecy mechanism that achieves perfect expected secrecy exploiting the channel model and the system dynamics. Later on, in Section \ref{Section_Relaxed} we explore the tradeoff between secrecy against the eavesdropper and utility to the user. 

\begin{remark}\label{remark:iid_model}
Modeling the user reception as an i.i.d. sequence implies a lossy memoryless channel, commonly assumed in networked control systems~\citep{hespanha2007survey} or information theory \citep{Cover2012elements}. The assumption that packet interception at the eavesdropper is also i.i.d. is novel, and can be similarly thought to model a lossy memoryless channel. In practical scenarios the eavesdropper cannot perfectly intercept all messages, e.g., it overhears the communication from a distance. Randomness in the interception may be attributed to random varying channel conditions of the wireless medium.\hfill $\diamond$
\end{remark}

\begin{remark}\label{REM_adversary_knowledge}
The assumption that system designer knows exactly the capabilities of the adversary, in our case the packet interception rate $p_2$ at the eavesdropper, may not be realistic. It is however a common assumption in formulations of security problems, for example, a channel model for the eavesdropper is assumed in the physical layer secrecy problem \citep{Proc_IEEE_Special_issue}. Alternatively the value $p_2$ can be thought of as a level of confidence the system designer has on the ability of an eavesdropper to intercept the messages or not. \hfill $\diamond$
\end{remark}

\section{Perfect Expected Secrecy}\label{Section_Perfect_Expected_Secrecy}

In this section, we explore sufficient conditions, under which, we can achieve perfect expected secrecy. In particular, we propose a simple mechanism, which  consists of flipping a coin with success probability $p$ at each time $k$ to decide whether to transmit the sensor's output measurement over the communication channel. If sent, and if the respective packet is not dropped, it also reaches the user and/or the eavesdropper. With probability $1-p$, on the other hand, no message is sent, hence, neither the user nor the eavesdropper receive any information. Intuitively, the proposed mechanism tries to achieve secrecy by randomly withholding sensor information.
By selecting this probability $p$, we can control, to some extent, the amount of the information availability in both channel outputs, though not independently.
Formally, the secrecy mechanism has the following form: 
\begin{equation}\label{EQN_mechanism_model}
\hat{y}\paren{k}=\left\{ \begin{aligned}&y(k) \text{ with prob. }p\\&\varepsilon \text{ with prob. }1-p\end{aligned}\right.,\;\forall k\ge 0.
\end{equation}

The packet drops in the channel outputs are considered independent of the decisions of the mechanism in \eqref{EQN_mechanism_model}. 

The next theorem states that a sufficient condition for perfect expected secrecy, is that the authorized output of the channel is more reliable than the unauthorized one, i.e. $p_{1}>p_{2}$. If this condition holds, then we can use the secrecy mechanism \eqref{EQN_mechanism_model} in order to satisfy \eqref{EQN_Definition_Perfect_Secrecy_1}, \eqref{EQN_Definition_Perfect_Secrecy_2}, by carefully selecting  values for the probability $p$.
\begin{thm}[Conditions for Perfect Secrecy]\label{THR_perfect_secrecy_pc}\hfill
Consider system \eqref{EQN_system} and the channel model in \eqref{EQN_channel_model}, under Assumption \ref{ASSUM_nominal_error}.
Perfect expected secrecy is achieved within the family of mechanisms \eqref{EQN_mechanism_model} if and only if
\begin{equation}
	p_{1}>p_{2}.
\end{equation}
In particular, there exists a probability $p_{c}\in \left[0,1\right)$ such that all probabilities $p$ satisfying
\begin{equation}\label{EQN_interval_of_perfect_secrecy}
\frac{p_{c}}{p_{1}}< p\le \min\set{\frac{p_{c}}{p_{2}},1} 
\end{equation} 
are exactly those, which guarantee perfect expected secrecy.
\hfill $\diamond$
\end{thm}

The condition $p_{1}>p_{2}$ is a reasonable requirement for secrecy in most cases of practical interest. For example, as mentioned in Remark~\ref{remark:iid_model}, when the eavesdropper intercepts the communication from some distance while the user is physically closer to the sensor, the user experiences better reception.

In the rest of this section, we present the proof of Theorem \ref{THR_perfect_secrecy_pc} and we characterize the probability $p_{c}$. 
Due to the channel model \eqref{EQN_channel_model} and the mechanism \eqref{EQN_mechanism_model}, at each time $k$, the user receives $y\paren{k}$ with probability $\hat{p}_{1}=pp_{1}$, and gets the no information symbol $\varepsilon$ with probability $1-\hat{p}_{1}$. Similarly, the eavesdropper receives $y\paren{k}$ with probability $\hat{p}_{2}=pp_{2}$, and $\varepsilon$ with probability $1-\hat{p}_{2}$.  Hence, the estimation problems of the user and the eavesdropper, are actually two separate estimation problems with intermittent observations \citep{Sinopoli2004Kalman}. Let $\gamma_{i}\paren{k}=1$ when $\hat{y}_{i}\paren{k}\neq \varepsilon$ and $\gamma_{i}\paren{k}=0$, when $\hat{y}_{i}\paren{k}= \varepsilon$. Then, the optimal estimates \eqref{eq:estimate_definition} follow the expressions for the intermittent Kalman filter given by:
\begin{equation}\label{EQN_Estimate}
\begin{split}
\hat{x}_{i}\paren{k+1}&=A\hat{x}_{i}\paren{k}+\gamma_{i}\paren{k}K_{i}\paren{k+1}\\&\times \paren{y\paren{k+1}-Cx\paren{k+1}}
\end{split}
\end{equation}
for $i=1,\,2$, where $K_{i}\paren{k}=P_{i}\paren{k}C^{\intercal}\paren{CP_{i}\paren{k}C^{\intercal}+R}^{-1}$ is the standard Kalman filter gain matrix. The prediction error \eqref{EQN_prediction_covariance} evolves recursively as:
\begin{equation}\label{EQN_Prediction_Error_Modified_Kalman}
P_{i}\paren{k+1}=g_{\gamma_{i}\paren{k}}\paren{P_{i}\paren{k}}
\end{equation}
where the function $g$ is defined as
\begin{equation}\label{EQN_function_g}
\begin{aligned}
g_{\lambda}\paren{X}&=AXA^{\intercal}+Q\\ &-\lambda A XC^{\intercal}{\paren{CXC^{\intercal}+R}^{-1}}CXA^{\intercal}
\end{aligned}
\end{equation}
for any $\lambda \in \clint{0,1}$ and any $X\succeq 0$ in $\REAL^{n\times n}$. Notice that in contrast to the classical Kalman filter, here $P_{i}\paren{k}$  is stochastic and depends on the random sequence $\gamma_{i}\paren{k}$ of successful receptions.

The estimation performance with intermittent observations varies as the probability of correct packet reception varies, in our case either $\hat{p}_{1}$ or $\hat{p}_{2}$. Specifically, there exists a critical probability $p_{c}$, which determines a kind of phase transition for the evolution of the expected prediction error covariance matrix $\expe{P_{i}\paren{k}}$. If $\hat{p}_{i}>p_{c}$, then the expected error covariance matrix is bounded. On the other hand, if $\hat{p}_{i}\le p_{c}$ then the expected error is unbounded. The following lemma extends results of \citep{Sinopoli2004Kalman} in our setup -- see Remark~\ref{remark:Sinopoli_difference} .

\begin{lem} \label{THR_critical_probability}
Given the system \eqref{EQN_system} and the secrecy mechanism \eqref{EQN_mechanism_model}, there exists a critical probability $p_{c}\in\left[0,1\right)$ such that:
\begin{align}
\label{EQN_lemma_diverging_error}
&\lim_{k\rightarrow \infty}\Tr\expe{P_{i}\paren{k}}=\infty\text{ if }\hat{p}_{i}\le p_{c},\: \forall \Sigma_{0}\succ 0
\\
\label{EQN_lemma_bounded_error}
&\sup\limits_{k\ge 0}\Tr\expe{P_{i}\paren{k}}\le M_{\Sigma_{0}},\text{ if } \hat{p}_{i}> p_{c},\: \forall \Sigma_{0}\succ 0
\end{align}
 where $i=1,\:2$ and $M_{\Sigma_{0}}$ is a positive constant, depending on the initial covariance $\Sigma_{0}$.\hfill $\diamond$
\end{lem}

\begin{pf}
The proof is included in the Appendix.
\end{pf}

Now we can take advantage of the phase transition according to Lemma \ref{THR_critical_probability} to chose mechanism \eqref{EQN_mechanism_model} for perfect secrecy, and thus prove Theorem \ref{THR_perfect_secrecy_pc}. First, we select $p$ to be small enough so that the eavesdropper's error $\Tr\expe{P_2\paren{k}}$ grows unbounded, so that condition \eqref{EQN_Definition_Perfect_Secrecy_1} of perfect expected secrecy is satisfied. According to \eqref{EQN_lemma_diverging_error} in Lemma \ref{THR_critical_probability}, this is guaranteed by selecting $\hat{p}_{2}=pp_{2}\le p_{c}$. On the other hand, $p$ should not be too small, so that the user's error $\Tr{\expe{P_1\paren{k}}}$ stays bounded and condition \eqref{EQN_Definition_Perfect_Secrecy_2} of perfect expected secrecy is satisfied. To achieve this, from condition \eqref{EQN_lemma_bounded_error} of Lemma \ref{THR_critical_probability}, we could select $\hat{p}_{1}=pp_{1}>p_{c}$. Combining both conditions, and due to the fact that $p\in\clint{0,1}$, it is sufficient select $p$ within the interval $p_{c}/p_{1}<p\le \min\set{p_{c}/p_{2},1}$. What remains to show is that this interval is nonempty. By the condition $p_{2}<p_{1}$, we obtain that $p_{c}/p_{1}<p_{c}/p_{2}$. It remains to argue that also $p_{c}/p_{1}<1$. By Assumption \ref{ASSUM_nominal_error}, the user's error is finite under no secrecy mechanism -- in our case when $p=1$. By \eqref{EQN_lemma_bounded_error} in Lemma \ref{THR_critical_probability}, this can occur only if $p_{1}>p_{c}$. This completes the sufficiency part of Theorem~\ref{THR_perfect_secrecy_pc}.

Now let us argue about the necessity part of Theorem~\ref{THR_perfect_secrecy_pc}. Both conditions $pp_{2}\le p_{c}<pp_{1}$ are necessary within the family of mechanisms \eqref{EQN_mechanism_model}. If $pp_{2}> p_{c}$ or $pp_{1}\le p_{c}$, then one of the conditions of perfect expected secrecy is violated, for according to Lemma \ref{THR_critical_probability} either the eavesdropper's error is bounded or the user's error grows unbounded. Hence, condition $p_{2}<p_{1}$ is necessary within the family of mechanisms \eqref{EQN_mechanism_model} and, furthermore, all probabilities that achieve perfect expected secrecy satisfy \eqref{EQN_interval_of_perfect_secrecy}.  This completes the proof of Theorem~\ref{THR_perfect_secrecy_pc}.

Theorem \ref{THR_perfect_secrecy_pc} describes that it is possible to achieve perfect secrecy by selecting the probability $p$ of our mechanism to lie within a specific interval, in particular $\left(p_{c}/p_{1},\min\set{p_{c}/p_{2},1} \right]$. However, this approach might still not be constructive. The reason is that computing the critical probability value $p_{c}$ is hard in general, apart from some special cases \citep{PlarreBullo2009kalman, MoSinopoli2008characterization}. In the following section, we explore the tradeoff between secrecy and utility via an optimization framework, and by this we also obtain computationally efficient methods to tune the probability $p$ of our secrecy mechanism.

\begin{remark}\label{remark:Sinopoli_difference} 
The presence of the eavesdropper makes the problem different than the one presented in \citep{Sinopoli2004Kalman}. The main goal in the analysis of the intermittent Kalman filter, has been to guarantee bounded error for a user. In contrast, in our work we also require unbounded error for the eavesdropper. Lemma \ref{THR_critical_probability} is an extension, as it sheds more light to the unboundedness case. In \citep{Sinopoli2004Kalman}, it was only shown that unboundedness occurs for some $\Sigma_{0}\succeq 0$, while we here prove that it occurs for all $\Sigma_{0}\succ 0$.
\end{remark}

\section{Tradeoff between secrecy and utility} \label{Section_Relaxed}

The notion of perfect secrecy, according to Definition~\ref{def:perfect_secrecy}, requires infinite estimation error at the eavesdropper. This might be too conservative in practice. At the same time, the definition requires that the legitimate user has a bounded estimation error, but this might be impractically large sometimes. An apparent tradeoff arises between the requirement for secrecy and the utility to the user. In this section, we seek to explore this tradeoff via an optimization framework. 

More specifically, consider our secrecy mechanism \eqref{EQN_mechanism_model}. If the sensor sends measurements more frequently, by increasing the probability $p$, the user maintains a better state estimate. At the same time, however, the eavesdropper is able to intercept more messages, hence, the level of secrecy declines. Conversely, secrecy is reinforced and the user's error deteriorates if the sensor withholds information at a higher rate. We are interested, then, in designing our mechanism \eqref{EQN_mechanism_model} to minimize the error at the user, as long as the eavesdropper error remains larger than some desired bound. This is an optimization problem of the form
\begin{equation}\label{EQN_practical_secrecy_optimization}
\begin{aligned}
&\underset{p \in[0,1]}{\text{minimize}}  &&\limsup_{k\rightarrow \infty} \Tr{\paren{\expe{P_{1}\paren{k}}}} \\
&\text{subject to} && \liminf_{k\rightarrow \infty} \Tr{\paren{\expe{P_{2}\paren{k}}}}\ge M
\end{aligned}
\end{equation}

where the design variable is the probability $p$ of mechanism \eqref{EQN_mechanism_model}. Here, $M>0$ is a positive constant describing some desired level of eavesdropper's error. This is a relaxed notion of secrecy as compared to the perfect secrecy in Definition~\ref{def:perfect_secrecy}. The latter can be recovered as $M \rightarrow \infty$.

Unfortunately, we cannot express the objective function or the constraint of problem \eqref{EQN_practical_secrecy_optimization} as a function of our secrecy mechanism \eqref{EQN_mechanism_model} in closed form. The reason is that, to the best of our knowledge, there are no closed form expressions for the expected error of the intermittent Kalman filter. Nonetheless, we can exploit well-known upper and lower bounds on the error introduced in \cite{Sinopoli2004Kalman}. In particular, we have the following two results.

\begin{prop} 
\label{THR_critical_prob_lower_bound}
Consider system \eqref{EQN_system} and secrecy mechanism \eqref{EQN_mechanism_model}. Let $p_{\ell}$ be a probability defined as follows
\begin{equation}\label{EQN_critical_prob_lower_bound}
p_{\ell}=1-1/\rho^{2}(A)
\end{equation}  
Then, the eavesdropper's error is asymptotically lower bounded by
\begin{equation}\label{EQN_covariance_prediction_error_limit_lower_bound}
\liminf_{k\rightarrow \infty} \Tr{\expe{P_{2}\paren{k}}}\ge \Tr{S\paren{p}}.
\end{equation}
If $pp_{2}>p_{\ell}$, $S\paren{p}$ is defined as the positive definite solution of
\begin{equation}\label{EQN_lower_bound_algebraic_eqn}
S\paren{p}=\paren{1-pp_{2}}AS\paren{p}A^{\intercal}+Q
\end{equation}
while if $pp_{2}\le p_{\ell}$, $\Tr S\paren{p}$ is defined  to be $\infty$. \hfill $\diamond$
\end{prop}

\begin{prop} 
\label{THR_critical_prob_upper_bound}
Consider system \eqref{EQN_system} and mechanism \eqref{EQN_mechanism_model}. Let $p_{u}$ be a probability defined as
\begin{equation}\label{EQN_critical_prob_upper_bound}
p_{u}=\inf\set{\lambda\in \clint{0,1}:\:\exists X\text{ with } X\succeq g_{\lambda}\paren{X}}
\end{equation}
where $g_{\lambda}\paren{X}$ is defined in \eqref{EQN_function_g}.
Then, the user's estimation error is asymptotically upper bounded by:
\begin{equation}\label{EQN_covariance_prediction_error_limit_upper_bound}
\limsup_{k\rightarrow \infty} \Tr{\expe{P_{1}\paren{k}}}\le \Tr{V\paren{p}}.
\end{equation}
If $pp_{1}>p_{u}$, $V\paren{p}$ is defined as the positive definite solution of:
\begin{equation}\label{EQN_upper_bound_algebraic_eqn}
V\paren{p}=g_{pp_{1}}\paren{V\paren{p}}, 
\end{equation}
while if $pp_{1}\le p_{u}$, $\Tr V\paren{p}$ is defined to be $\infty$. \hfill $\diamond$
\end{prop}

Propositions \ref{THR_critical_prob_lower_bound}, \ref{THR_critical_prob_upper_bound} follow from the proofs of Theorems 3, 4 in \citep{Sinopoli2004Kalman}.

We can utilize the bounds of the preceding propositions to design a desired secrecy mechanism, by relaxing the optimization problem \eqref{EQN_practical_secrecy_optimization}. In particular, we relax the constraint of \eqref{EQN_practical_secrecy_optimization} by requiring that the lower bound $S\paren{p}$ on the eavesdropper's estimation error is larger than the desired value $M$. Moreover we relax the objective with the upper bound $V\paren{p}$ on the user's estimation error. Thus, the relaxation of problem \eqref{EQN_practical_secrecy_optimization} has the form:
\begin{equation}\label{EQN_practical_secrecy_replace_bounds_optimization}
\begin{aligned}
&\underset{p \in[0,1]}{\text{minimize}}  &&\Tr{V\paren{p}} \\
&\text{subject to} && \Tr{S\paren{p}}\ge M
\end{aligned}
\end{equation}

Even though Problem \eqref{EQN_practical_secrecy_replace_bounds_optimization} is not convex, it has a specific structure 
that allows us to solve it efficiently. In particular, both the objective value $ \Tr S\paren{p}$ and the constraint $\Tr V\paren{p}$ are monotonically decreasing functions with respect to $p$ (see also Lemma \ref{LEM_monotonicity} in the Appendix). The next theorem hence explicitly describes the optimal solution.

\begin{thm}\label{THR_optimal_solution_bounds_optimization_problem}
Consider system \eqref{EQN_system} and the mechanism \eqref{EQN_mechanism_model}. Let $\Tr{S\paren{p}}$, $\Tr{V\paren{p}}$ be the lower and upper bounds, as defined in Proposition \ref{THR_critical_prob_lower_bound} and Proposition \ref{THR_critical_prob_upper_bound}  respectively. Then, the optimal solution of problem \eqref{EQN_practical_secrecy_replace_bounds_optimization} is given by
\begin{equation}\label{EQN_p_star}
	p^\star= \max\{ p \in [ 0,1]: \, \Tr{S\paren{p}} \geq M\}.	
\end{equation}
\end{thm}
\begin{pf}
In Lemma \ref{LEM_monotonicity}, included in the Appendix, we proved that $\Tr S\paren{p}$, $\Tr V\paren{p}$ are non-increasing functions of $p$. As a result, problem \eqref{EQN_practical_secrecy_replace_bounds_optimization} is equivalent to the optimization problem $\max\{ p \in [0,1]: \, \Tr{S_{2}\paren{p}} \geq M\}$. \hfill $\qed$
\end{pf}

Capitalizing on the above result and monotonicity, we may further devise an algorithm to find the optimal solution $p^*$ of problem \eqref{EQN_practical_secrecy_replace_bounds_optimization}, based on a bisection search.
This process is presented in Algorithm \ref{ALG_solution_practical_with_bounds}. It takes as inputs the system and noise parameters $A$, $C$, $Q$, $R$, probability $p_{2}$, the desired bound $M>0$ on the eavesdropper's error and a positive constant $\varepsilon$. This constant $\varepsilon$ represents the absolute tolerance within which we want to compute $p^{\star}$. 

The steps of the algorithm are the following. First, the probability $p_{\ell}$ is computed, which according to \eqref{EQN_critical_prob_lower_bound}, depends only on the spectral radius of matrix $A$, and, thus, can be readily computed. 
Then, a bisection on the interval $p \in [0, 1]$ is performed in order to solve \eqref{EQN_p_star}. 
 At each bisection step, the algorithm evaluates the function $\Tr{S(p)}$ at the midpoint $p$ of the current interval. If $pp_{2}\le p_{l}$, then the algorithm sets $\Tr{S(p)}=\infty$. Otherwise, the algorithm solves the linear matrix equality \eqref{EQN_lower_bound_algebraic_eqn} with respect to  $S(p)$ for the given $p$, e.g., solved as a Lyapunov equation.
Then, the computed value $\Tr{S(p)}$ is compared with the desired value $M$, and the half-interval for the next bisection step is selected, based on the fact that $\Tr{S(p)}$ is decreasing in $p$.
Algorithm \ref{ALG_solution_practical_with_bounds} terminates when $p^{\star}$ is known to lie within an interval of $\varepsilon$ tolerance. The algorithm terminates after at most $-\log_{2}\paren{\varepsilon}$ iterations, since at each iteration, the bisection search ends up with half of the interval of the previous step.

\begin{algorithm}
 \caption{ Optimal probability $p^\star$ of problem \eqref{EQN_practical_secrecy_replace_bounds_optimization}}
 \label{ALG_solution_practical_with_bounds}
 \begin{algorithmic}[1] {}
 \Require $A$, $C$, $Q$, $R$, $p_{2}$, $M$, $\epsilon$
 \Ensure optimal solution $p^\star$
 \State{Compute $p_{\ell}$. }  
 \State{Do bisection on $p$. Values $u$ and $l$ are the upper and lower bounds in every step of the bisection.}
	\State{Set $\ell=0$, $u=1$ for the initial bounds.} 
	 \While{$\abs{u-\ell}<\epsilon$}
	 \State{Set $p=\paren{\ell+u}/2$}
	 	\State{Set $\Tr S\paren{p}=\infty$ if $pp_{2}\le p_{l}$. Else, compute $\Tr S\paren{p}$. }
		\If{$\Tr{S\paren{p}}<M$}
			\State{Set $u=p$} 
		\Else
			\State{Set $l=p$}
		\EndIf
 	\EndWhile
 \State{\Return $p$}
 \end{algorithmic}
\end{algorithm}

Finally, we point out that after Algorithm \ref{ALG_solution_practical_with_bounds} returns the optimal probability $p^\star$, we may also evaluate the optimal objective value $\Tr V\paren{p^\star}$ of \eqref{EQN_practical_secrecy_replace_bounds_optimization}. As long as $p^\star p_{1}> p_{u}$ this optimal value is finite and can be computed by solving \eqref{EQN_upper_bound_algebraic_eqn} with respect to $V\paren{p^\star}$. The value of probability $p_{u}$ can be computed by a quasi-convex optimization problem, while equation  \eqref{EQN_upper_bound_algebraic_eqn} can be solved by a semidefinite program (see \cite{Sinopoli2004Kalman} for both methods).

\section{Examples}\label{sec:scalar}

In this section, we present analytical expressions for the problem of secrecy in the special case of scalar systems, as well as numerical results for a second order system. In the former case, we are interested in the tradeoff between secrecy and utility to the user, as captured by the mechanisms $p^*$ that solve problem \eqref{EQN_practical_secrecy_replace_bounds_optimization}. In particular, we examine how the solution depends on the system's and channel's parameters and how the solution varies, as the required bound $M$ on the eavesdropper error varies.  

Here all parameters $A,\,C,\,R,\,Q,\,\Sigma_{0}\in \REAL$ as well as the upper bound $V\paren{p}\in \REAL$ on the user's error and the lower bound $S\paren{p}\in \REAL$ on the eavesdropper's error, as defined in \eqref{EQN_lower_bound_algebraic_eqn} and \eqref{EQN_upper_bound_algebraic_eqn} respectively, are scalars. Moreover, the scalar system is one of the special cases where we know exactly the critical probability and $p_{\ell}=p_{u}=p_{c}=1-1/A^2$ \citep{Sinopoli2004Kalman}. 
The interesting case to study is when secrecy is needed, i.e. when the eavesdropper has bounded expected error without any secrecy mechanism ($p=1$). According to Lemma \ref{THR_critical_probability}, this happens when $p_{2}>p_{c}=1-1/A^2$. Also, since $S\paren{p}\ge S\paren{1}$, the secrecy constraint $S\paren{p}>M$ with $M< S\paren{1}$ is trivially satisfied. Thus, we also assume that $M\ge S\paren{1}$ in this section.

From \eqref{EQN_covariance_prediction_error_limit_lower_bound}, we can find that $S\paren{p}=Q/\paren{1-(1-pp_{2})A^2}$. Thus, by Theorem \ref{THR_optimal_solution_bounds_optimization_problem}, we deduce that the optimal secrecy mechanism $p^\star$ that solves problem \eqref{EQN_practical_secrecy_replace_bounds_optimization} is the probability value that satisfies $S\paren{p^\star}=M$. As we can see, for the scalar case, we can obtain the following closed form expression for $p^*$:
\begin{equation}
	p^\star=\frac{p_{c}}{p_{2}}+\frac{Q}{Mp_{2}A^2}.
\end{equation}
Observe that the first term $\frac{p_{c}}{p_{2}}$, on the right hand side, is a probability that guarantees perfect expected secrecy if $p_{1}>p_{2}$ (see the discussion of Section \ref{Section_Perfect_Expected_Secrecy}). Hence as $M\rightarrow \infty$, we may recover perfect expected secrecy.

Next, we evaluate the upper bound on the user's error when using the above mechanism $p^*$, i.e., the optimal objective value of problem \eqref{EQN_practical_secrecy_replace_bounds_optimization}. Recall that if $p^\star p_{1}\le p_{c}$, then $V\paren{p^\star}=\infty$. Otherwise, if $p^\star p_{1}>p_{c}$, then we can solve for the positive solution $V\paren{p^\star}$ of the (quadratic) equation \eqref{EQN_upper_bound_algebraic_eqn}. So, in the latter case, we obtain:

\begin{equation*}
V\paren{p^\star}=\frac{\beta+\sqrt{\beta^{2}+4QRC^2\left[\paren{A^2-1}\paren{\frac{p_{1}}{p_{2}}-1}+\frac{p_{1}}{M p_{2}}\right]}}{2C^2\left[\paren{A^2-1}\paren{\frac{p_{1}}{p_{2}}-1}+\frac{p_{1}}{M p_{2}}\right]}
\end{equation*}
where $\beta=\paren{A^2-1}R+Q^2C$. 

 \begin{figure}[t] \centering{
\includegraphics[scale=0.45]{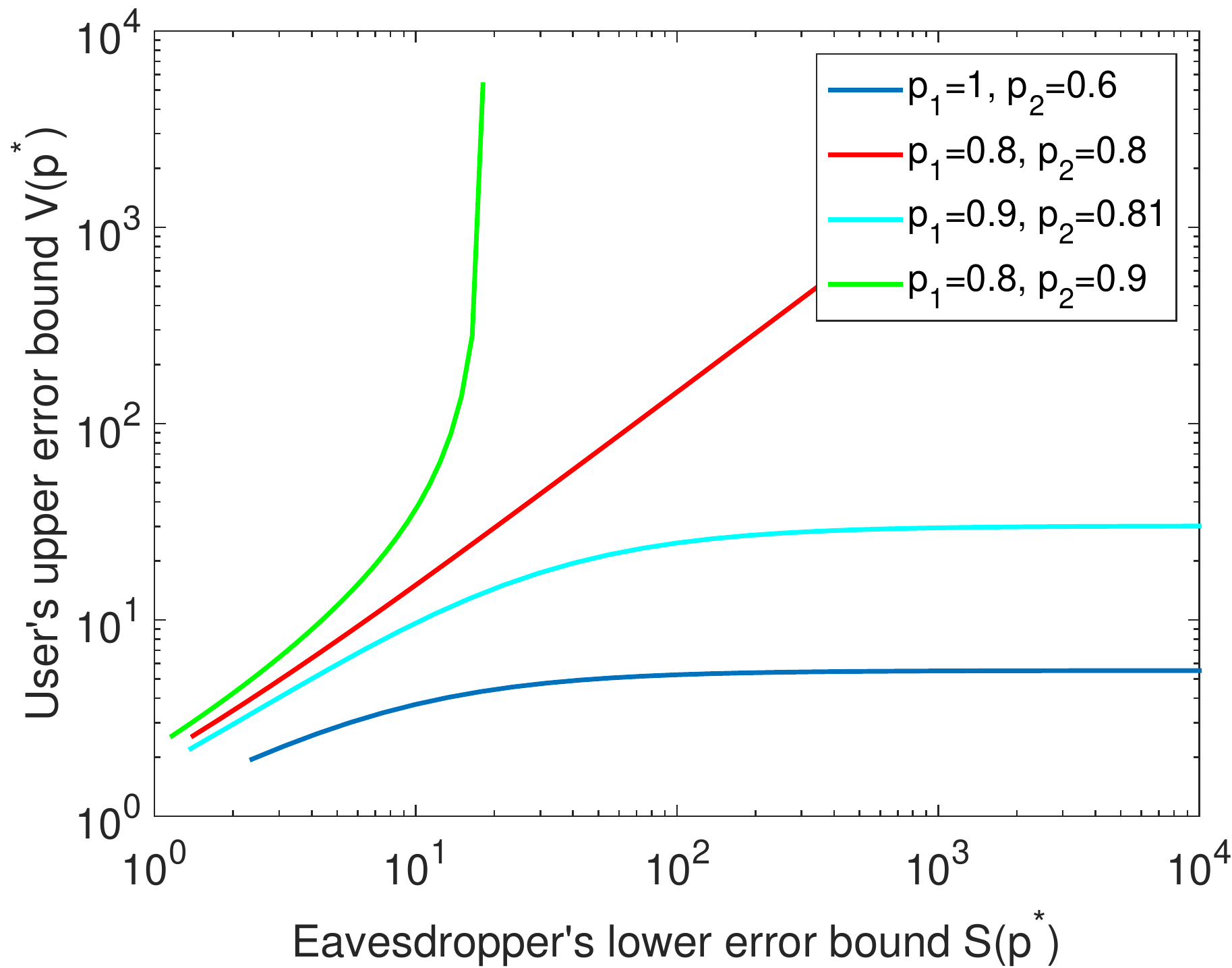}}   
\caption { This figure shows different tradeoffs between optimum user's upper bound $V\paren{p^\star}$ and eavesdropper's lower bound $S\paren{p^\star}$ in problem \eqref{EQN_practical_secrecy_replace_bounds_optimization}, as we vary $p_1$, $p_2$. As the ratio $p_1/p_2$ increases, tradeoff between secrecy and estimation efficiency improves. As we observe,  tradeoff is in favour of the user when $p_{1}>p_{2}$, but in favor of the eavesdropper when $p_{2}>p_{1}$.}
\label{Figure_UpperVsLower}
\end{figure}

The expression for $V\paren{p^\star}$ as a function of $M$ captures the tradeoff between the guaranteed utility to the user and the secrecy level $M$ at the eavesdropper. Figure \ref{Figure_UpperVsLower} plots this expression as a function of $M$ for different values of the channel probabilities $p_{1}$, $p_{2}$. The system parameters were $A=1.2$, $C=1$, $Q=1$ and $R=1$. Interestingly, as the ratio $p_1/p_2$ increases, the tradeoff between secrecy and efficiency improves, meaning that a better estimation error can be guaranteed at the user at a given secrecy level. 
 \begin{figure}[b] \centering{
\includegraphics[scale=0.45]{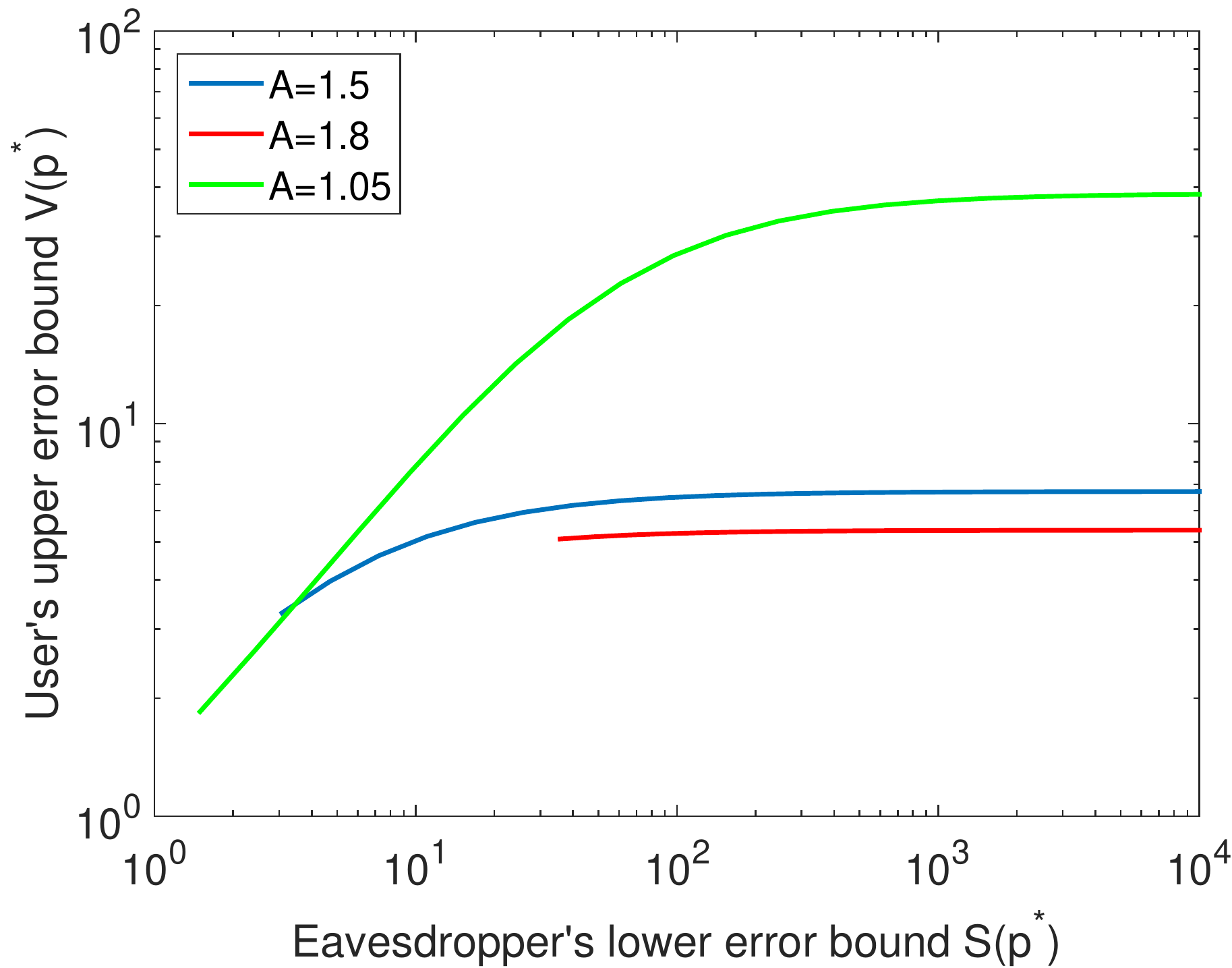}}   
\caption { his figure shows different tradeoffs when we vary $A$. Here we fix probabilities $p_{1}=0.9$ and $p_{2}=0.7$. As $A$ increases, tradeoff between secrecy and estimation efficiency improves.}
\label{Figure_Effect_of_A}
\end{figure}
In Figure \ref{Figure_Effect_of_A}, we also capture the effect of the dynamics in the secrecy-utility tradeoff, for $C=1$, $Q=1$, $R=1$ and $p_{1}=0.9$, $p_{2}=0.7$. When the unstable eigenvalue grows, the tradeoff improves.

Until now we have only explored how the user and eavesdropper errors behave in expectation and in the limit per Definition~\ref{def:perfect_secrecy}. Let us now present one random time sample of the behaviour of the actual estimation errors $\norm{\hat{x}_{i}\paren{k}-x\paren{k}}_{2}$, for user and eavesdropper $i=1$, $2$ respectively. We simulate a second order system, along with the state estimate given by \eqref{EQN_Estimate},  with $A=\matr{{cc}1.2&1\\0& 1.1}$, $C=\matr{{cc} 1& 0}$, $R=1$, $Q=\Sigma_{0}=\matr{{cc}1&0.5\\0.5& 2}$ and probabilities $p_{1}=0.9$, $p_{2}=0.6$. 
We compare the estimation errors between the user and the eavesdropper for two cases; i) when no secrecy mechanism is employed ($p=1$), ii) using the secrecy mechanism with probability $p=p_\ell/p_2=0.51$. To keep the comparison fair, we use the same randomly generated sample for the noise and packet drop sequences, in both cases. 
\begin{figure}[t] \centering{
\includegraphics[scale=0.5]{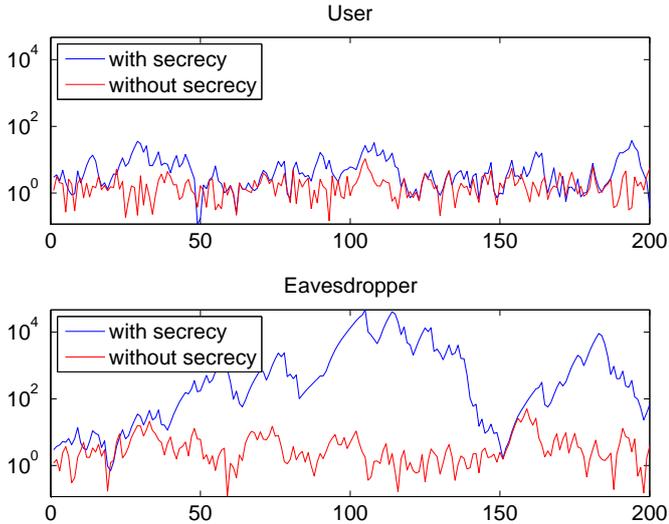}}   
\caption {This figure shows how the estimation errors $\norm{\hat{x}_{1}\paren{k}-x\paren{k}}_{2}$, $\norm{\hat{x}_{2}\paren{k}-x\paren{k}}_{2}$ evolve over time for two cases. In the first case, no secrecy mechanism is employed ($p=1$). In the second case, a secrecy mechanism with $p=0.51$ is used. The same randomly generated sequence of noises and packet drops was used in both cases. The packet drops of the user are independent of the packet drops of the eavesdropper.}
\label{Figure_TimeTraceEstimation1}
\end{figure}

In Figure \ref{Figure_TimeTraceEstimation1} we see that there is benefit in employing the secrecy mechanism. When no secrecy is used, we observe that the user's and eavesdropper's errors are comparable, and the confidentiality of the system's state is compromised.  The time average $\frac{1}{N}\sum_{k=1}^{N}\norm{\hat{x}_{i}\paren{k}-x\paren{k}}_{2}$ of the errors, for $N=200$, are $1.9$ for the user and $4.6$ for the eavesdropper respectively.  However, in the later case, where a secrecy mechanism with  $p=0.51$ is used, the relative gain we have improves. The time average of the absolute  errors are $6.2$ for the user and $ 2922.7$ for the eavesdropper respectively.

We close with a final observation that is not evident from considering only the prediction errors in expectation per Definition~\ref{def:perfect_secrecy}. In particular, the eavesdropper's error drops to small values infinitely often, as one can see in Figure \ref{Figure_TimeTraceEstimation1}. This occurs because whenever the eavesdropper successfully intercepts a measurement, its prediction covariance matrix 
decreases dramatically -- see also \cite{WuJohansson2014probabilistic} for a related discussion. In other words, the fact that the eavesdropper's error $\Tr \expe{P_{2}\paren{k}}$  grows unbounded in expectation, by our secrecy definition, does not imply that the actual  sample $\Tr P_{2}\paren{k}$ is always large.

\section{Conclusion}\label{sec:conclusion}
In this paper, we considered the novel problem of remote estimation in the presence of an eavesdropper.
Requiring confidentiality makes the problem challenging, and forces us to modify the communication scheme, by adding a secrecy mechanism to the sensor. 
A simple secrecy mechanism, which randomly withholds measurements, can guarantee perfect expected secrecy, 
if the user's reception ability is better than the eavesdropper's.
This mechanism creates a tradeoff between confidentiality and utility, which can be approximated by solving an optimization problem. The numerical results of the second order system show that our proposed mechanism can be useful in practice. Still, it guarantees secrecy in expectation -- with nonzero probability the eavesdropper's uncertainty can be small. 
In future work, we will address more general channel models and alternative mechanisms for higher performance.
We will also study whether the condition, that the user's rate is higher than the eavesdropper's, is necessary for general mechanisms.
Other extensions include the case where the sensor receives packet acknowledgements from the user.
\bibliography{Paper_wireless_secrecy}             

\appendix
\section{Proof of results}    
\subsection*{Proof of Lemma \ref{THR_critical_probability}}
Since $\paren{A,C}$ is detectable, $\paren{A,Q^{\frac{1}{2}}}$ is controllable and $A$ is unstable, the result of Theorem 2 in \citep{Sinopoli2004Kalman} readily applies. By this result, there exists a $p_{c}\in \left[0,1\right)$, such that 
\begin{align}
&\exists \Sigma_{0}\succeq 0:\:\lim_{k\rightarrow \infty}\Tr\expe{P_{i}\paren{k}}=\infty\text{, if }\hat{p}_{i}\le p_{c} 
\label{EQN_Sinopoli_unbounded}
\\
&\sup\limits_{k\ge 0}\Tr\expe{P_{i}\paren{k}}\le M_{\Sigma_{0}},\text{ if } \hat{p}_{i}> p_{c},\: \forall \Sigma_{0}\succeq 0
\end{align}
 where $i=1,\:2$ and $M_{\Sigma_{0}}$ is a positive constant, depending on the initial covariance $\Sigma_{0}$. This result directly implies the statement \eqref{EQN_lemma_bounded_error}. Hence it remains to show that \eqref{EQN_lemma_diverging_error} holds for any positive definite initial condition $\Sigma_{0}\succ 0$.

Let $\Sigma_{0}^\prime\succeq 0$ be one initial condition with $P_{i}^\prime\paren{k}$ the respective error sequence, for which $\lim_{k\rightarrow \infty}\Tr\expe{P_{i}^\prime\paren{k}}=\infty$, according to \eqref{EQN_Sinopoli_unbounded}.
Also, let $P_{i}\paren{k}$ be the error sequence with some arbitrary initial condition $\Sigma_{0}\succ 0$. Both covariance errors $P_{i}^\prime\paren{k}$ and $P_{i}\paren{k}, k \geq 0$ are random variables that depend on the sequence of packet successes $\gamma_i(k), k \geq 0$ according to \eqref{EQN_Prediction_Error_Modified_Kalman}.
To complete the proof, it is sufficient to find a positive constant $\beta>1$ such that $\Tr\expe{P_{i}^\prime\paren{k}}\le \beta \Tr\expe{P_{i}\paren{k}}$ for all $k\ge 0$. Then, in the limit $\lim_{k\rightarrow \infty}\Tr\expe{P_{i}\paren{k}}\ge \paren{1/\beta}\lim_{k\rightarrow \infty}\Tr\expe{P_{i}^\prime\paren{k}}=\infty$ and the result \eqref{EQN_lemma_diverging_error} follows. 
Note that as $\Sigma_{0}\succ 0$, we can find a large enough positive constant $\beta>1$ such that $\beta\Sigma_{0}\succeq \Sigma^\prime_{0}$. Then, we claim that for any fixed packet success sequence $\gamma_{i}\paren{k}$, $k \geq 0$, we have $P_{i}^\prime\paren{k}\preceq \beta P_{i}\paren{k}$ for all $k\ge 0$. Hence, averaging over all possible success sequences, we obtain $\Tr\expe{P_{i}^\prime\paren{k}}\le \beta \Tr\expe{P_{i}\paren{k}}$.

Finally, to prove that $P_{i}^\prime\paren{k}\preceq \beta P_{i}\paren{k}$ for all $k\ge 0$, we use induction. At $k=0$ we have $\Sigma^\prime_{0}\preceq\beta\Sigma_{0}$, which is the same as $P_{i}^\prime\paren{0}\preceq \beta P_{i}\paren{0}$. Suppose that $P_{i}^\prime\paren{k}\preceq \beta P_{i}\paren{k}$. Then, at $k+1$ we employ the relation \eqref{EQN_Prediction_Error_Modified_Kalman}- \eqref{EQN_function_g}. Notice that in \eqref{EQN_function_g}, $g_{\lambda}\paren{X}$ is an increasing function of $X$ with respect to the positive semidefinite cone for any $\lambda \in\clint{0,1}$ (see Lemma 1 in \cite{Sinopoli2004Kalman}). We can also see that $g_{\lambda}\paren{X}$ is an increasing function of $Q$ and $R$ as well. Hence, for $\beta>1$, since $\beta Q\succ Q$, $\beta R \succ R$, we obtain, that $\beta g_{\lambda}\paren{X}\succeq g_{\lambda}\paren{\beta X}$. 
Then, at time $k$,  from \eqref{EQN_Prediction_Error_Modified_Kalman}, we have
\begin{align*}
\beta P_{i}\paren{k+1}=&\beta g_{\gamma_{i}\paren{k}}\paren{P_{i}\paren{k}}\\
\succeq & g_{\gamma_{i}\paren{k}}\paren{\beta P_{i}\paren{k}}
\\\succeq & g_{\gamma_{i}\paren{k}}\paren{P_{i}^\prime\paren{k}}=P^\prime_{i}\paren{k+1}. 
\end{align*}
The first inequality comes from the fact that $\beta>1$, while the second comes from monotonicity of $g_{\lambda}\paren{X}$ with respect to $X$and the induction hypothesis at time step $k-1$. \hfill $\qed$

\begin{lem} \label{LEM_monotonicity}
The bounds $\Tr S\paren{p}$, $\Tr V\paren{p}$, defined in \eqref{EQN_lower_bound_algebraic_eqn} and \eqref{EQN_upper_bound_algebraic_eqn}, defined to be $\infty$ when $pp_{2}\le p_{l}$ and $pp_{1}\le p_{u}$ respectively, are non-increasing functions of $p$.
\end{lem}
\begin{pf}
First, we prove that $\Tr S\paren{p}$ is non-increasing with $p$. From the proof of Theorem 3 in \citep{Sinopoli2004Kalman}, we know that when $pp_{2}>p_{l}$, then $S\paren{p}=\lim_{k\rightarrow  \infty }S_{k+1}\paren{p}$, is the limit of a sequence of matrices $S_{k+1}\paren{p}=m_{p}\paren{S_{k}\paren{p}}$, where $m_{p}\paren{X}=\paren{1-pp_{2}}AXA^\intercal+Q$ and $S_{0}\paren{p}=0$. Now observe that $m_{p}\paren{X}$ is decreasing with $p$ and increasing with $X$ (with respect to the positive semidefinite cone). Given two different probabilities  $1\ge\lambda_{1}\ge\lambda_{2}>p_{l}/p_{2}$, we will use the monotonicity of $m_{p}\paren{X}$ to show by induction that $S_{k}\paren{\lambda_{1}}\preceq S_{k}\paren{\lambda_{2}}$, for all $k\ge 0$. For $k=0$, it is true, since $S_{0}\paren{\lambda_{i}}=0$, $i=1$, $2$. Now, assume that $S_{k-1}\paren{\lambda_{1}}\preceq S_{k-1}\paren{\lambda_{2}}$ holds. Then, 
\begin{align*}
S_{k}\paren{\lambda_{1}}&=m_{\lambda_{1}}\paren{S_{k-1}\paren{\lambda_{1}}}\\ &\preceq m_{\lambda_{2}}\paren{S_{k-1}\paren{\lambda_{1}}}\\&\preceq m_{\lambda_{2}}\paren{S_{k-1}\paren{\lambda_{2}}}=S_{k}\paren{\lambda_{2}}
\end{align*}
where the first inequality comes from monotonicity of $m_{p}\paren{X}$ with respect to $p$ and $\lambda_{1}\ge \lambda_{2}$;  the second inequality comes from monotonicity with respect to $X$ and the induction hypothesis. 
Taking the trace in both sides we have that $\Tr S_{k}\paren{\lambda_{1}}\le \Tr S_{k}\paren{\lambda_{2}}$ and as $k\rightarrow \infty$, we obtain $\Tr S\paren{\lambda_{1}}\le \Tr S\paren{\lambda_{2}}$.
Since $\Tr S\paren{p}$ is extended to $\infty$ for $pp_{2}\le p_{l}$,  $\Tr S\paren{p}$ is non-increasing in all of $\clint{0,1}$.

The proof that $\Tr V\paren{p}$ is non-increasing with $p$ is similar. From the proof of Theorem 4 in \cite{Sinopoli2004Kalman}, we have that when $pp_{1}>p_{u}$, then $V\paren{p}$ is the limit of a sequence of matrices $V_{k+1}\paren{p}=g_{pp_{1}}\paren{V_{k}\paren{p}}$, with $V_{0}\paren{p}=\Sigma_{0}$. Function $g_{pp_{1}}\paren{X}$ is defined in Proposition \ref{THR_critical_prob_upper_bound} and it is also decreasing with respect to $p$ and increasing with respect to $X$ (see Lemma 1 in \citep{Sinopoli2004Kalman}). Repeating exactly the same argument as before, we can show that if $1\ge \lambda_{1}\ge \lambda_{2}>p_{u}/p_{1}$, then $ V_{k}\paren{\lambda_{1}}\preceq  V_{k}\paren{\lambda_{2}}$, for any $k\ge 0$. Therefore, as $k\rightarrow \infty$, we obtain $\Tr V\paren{\lambda_{1}}\le \Tr V\paren{\lambda_{2}}$. Since $V\paren{p}$ is extended to $\infty$ for $pp_{1}\le p_{u}$, then $V\paren{p}$ is non-increasing in all of $\clint{0,1}$.\hfill $\qed$
\end{pf}

\end{document}